\documentclass[aps,preprint,showpacs,showkeys]{revtex4}
\usepackage{graphicx}
\begin{document}
\preprint{Aswartham et al, Ba$_{1-x}$Na$_x$Fe$_2$As$_2$}
\title{Hole-doping in BaFe$_2$As$_2$: The case of Ba$_{1-x}$Na$_x$Fe$_2$As$_2$ single crystals}
\author{S.~Aswartham} 
\affiliation{Leibniz Institute for Solid State and Materials Research, IFW, D-01069 Dresden, Germany}
\author{M.~Abdel-Hafiez}
\affiliation{Leibniz Institute for Solid State and Materials Research, IFW, D-01069 Dresden, Germany}
\author{D.~Bombor}
\affiliation{Leibniz Institute for Solid State and Materials Research, IFW, D-01069 Dresden, Germany}
\author{M.~Kumar}
\affiliation{Leibniz Institute for Solid State and Materials Research, IFW, D-01069 Dresden, Germany}
\author{A.~U.~B.~Wolter}
\affiliation{Leibniz Institute for Solid State and Materials Research, IFW, D-01069 Dresden, Germany}
\author{C.~Hess}
\affiliation{Leibniz Institute for Solid State and Materials Research, IFW, D-01069 Dresden, Germany}
\author{D.~V.~Evtushinsky}
\affiliation{Leibniz Institute for Solid State and Materials Research, IFW, D-01069 Dresden, Germany}
\author{V.~B.~Zabolotnyy}
\affiliation{Leibniz Institute for Solid State and Materials Research, IFW, D-01069 Dresden, Germany}
\author{A.~A.~Kordyuk}
\affiliation{Leibniz Institute for Solid State and Materials Research, IFW, D-01069 Dresden, Germany}
\author{T.~K.~Kim}
\affiliation{Leibniz Institute for Solid State and Materials Research, IFW, D-01069 Dresden, Germany}
\author{S.~V.~Borisenko}
\affiliation{Leibniz Institute for Solid State and Materials Research, IFW, D-01069 Dresden, Germany}
\author{G.~Behr}
\affiliation{Leibniz Institute for Solid State and Materials Research, IFW, D-01069 Dresden, Germany}
\author{B.~B\"uchner}
\affiliation{Leibniz Institute for Solid State and Materials Research, IFW, D-01069 Dresden, Germany}
\author{S.~Wurmehl}
\email{s.wurmehl@ifw-dresden.de}
\affiliation{Leibniz Institute for Solid State and Materials Research, IFW, D-01069 Dresden, Germany}


\date{\today}

\begin{abstract}
Single crystals of Ba$_{1-x}$Na$_x$Fe$_2$As$_2$ with $x$ = 0, 0.25, 0.35, 0.4 were grown using a self-flux high temperature solution growth technique. The superconducting and normal state properties were studied by temperature dependent magnetic susceptibility, electrical resistivity and specific heat revealing that the magnetic and structural transition is rapidly suppressed upon Na-substitution at the Ba-site in BaFe$_2$As$_2$, giving rise to superconductivity. A superconducting transition as high as 34~K is reached for a Na-content of $x$=0.4. The positive Hall coefficient confirms that the substitution of Ba by Na results in hole-doping similarly to the substitution of Ba by K. Angle resolved photoemission spectroscopy was performed on all Ba$_{1-x}$Na$_x$Fe$_2$As$_2$ crystals. The Fermi surface of hole-doped Ba$_{1-x}$Na$_x$Fe$_2$As$_2$ is to high extent the same as the Fermi surface found for the K-doped sister compounds, suggesting a similar impact of the substitution of Ba by either K or Na on the electronic band dispersion at the Fermi level.

\end{abstract}

\maketitle

\section{Introduction}
The recent discovery of superconductivity in LaO$_{1-x}$F$_x$FeAs with critical temperatures up to 26~K \cite{KWH08} has significantly increased research activities in the field of superconductivity. Quickly after the first report on pnictide superconductors, the critical transition temperature $T_c$ has been enhanced to 55~K in SmO$_{1-x}$F$_x$FeAs \cite{RLY08}. Availability of different types of materials in this new family of superconductors allows addressing many open questions in superconductivity e.g.\ the question whether magnetism and superconductivity are co-existing or competing. The Fe-based superconductors are classified in to four types according to their stoichiometry, such as the series of 1111 ($R$OFeAs and $AE$FFeAs) \cite{KWH08,TJW08}, 122 ($AE$Fe$_2$As$_2$ and $A$Fe$_2$As$_2$) \cite{RTJ08,Sll08}, 111 (AFeAs) \cite{TTL08} and 11 (FeSe$_{1-x}$Te$_x$)\cite{HLK08}. Although the $T_c$ of the 1111 family is found to be the highest among the Fe-based superconductors, their intrinsic properties are yet unexplored. For instance, the detailed electronic structure is still not investigated, primarily due challenging requirements posed by the high pressure grows technique and, hence, due to the lack of large single crystals \cite{ZKB08}. Crystals of the 122 series are comparably easier to grow at ambient pressure and are more convenient to handle than e.g.\ the 111 compounds. Therefore the 122 compounds are the best studied pnictide superconductors.

One of the first publications on 122 compounds dates back to 1964 \cite{BS65}. Since then more than 700 compounds have been found to adopt the 122 structure type \cite{VC91}. All members of the 122 family constitute of the ThCr$_2$Si$_2$ structure-type and are sensitive to chemical substitutions and pressure (see e.g. \cite{CBNC09}). Usually, the parent compound is an antiferromagnet with spin density wave ordering in orthorhombic symmetry. Superconductivity can be achieved either by external pressure or by chemical substitution which leads to a tetragonal symmetry and supression of the magnetic ordering. Rotter $et$ $al.$ were the first to report superconductivity in K-doped BaFe$_2$As$_2$ with the highest $T_c$ of 38~K in the 122 series \cite{RTJ08}. However, the quest for new superconducting materials goes on and many new superconducting materials with different critical temperatures have been synthesized within the 122 series. The report of superconductivity in polycrystalline samples of Na-doped BaFe$_2$As$_2$ by Cortes-Gil $et$ $al.$ can be considered as a recent example \cite{CRD10}. Although, this study on polycrystals suggests that Ba$_{1-x}$Na$_x$Fe$_2$As$_2$ is quite similar to the corresponding K-substituted sister compounds, a detailed study of the physical properties is lacking in ref. \cite{CRD10} and in particular no single crystals were reported so far. It is, however, necessary to investigate in detail how physical properties vary with different dopants and doping levels, considering the fact that also the electronic structure is significantly modifed depending on the nature of the doping elements and the amount of doping. Here, extending the previous work, we report on the growth of Ba$_{1-x}$Na$_x$Fe$_2$As$_2$ single crystals with $x$ = 0, 0.25, 0.35, 0.4 and on their magnetic, electronic transport, thermodynamic and electronic properties.

\section{Materials and methods}
\begin{figure}
\centering
\includegraphics[width=8cm]{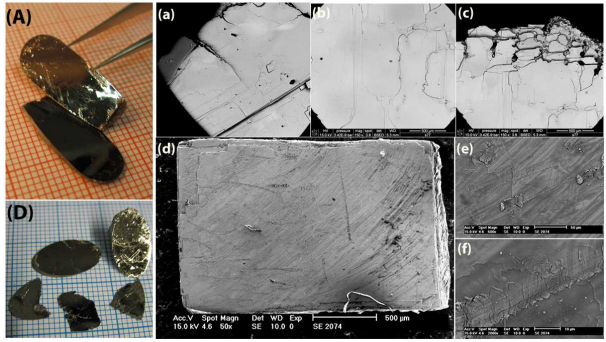}
\caption{ (A) Plate-like as grown single crystals of Ba$_{0.75}$Na$_{0.25}$Fe$_2$As$_2$ from self-flux, (a-c) SEM pictures from the same batch of  a Ba$_{0.75}$Na$_{0.25}$Fe$_2$As$_2$ single crystal.
(D) Plate-like as grown single crystals of Ba$_{0.65}$Na$_{0.35}$Fe$_2$As$_2$, (d-e) SEM pictures from the same batch of a Ba$_{0.65}$Na$_{0.35}$Fe$_2$As$_2$ single crystal which shows the typical layer by layer growth. }
\label{Fig_1}
\end{figure}

All preparation steps like weighing, mixing, grinding and storage were carried out in an Ar-filled glove-box (O$_2$ and H$_2$O level less than 0.1~ppm).
All precursor materials were prepared by reacting with As yielding e.g.\ BaAs, FeAs, and Fe$_2$As. These precursor materials were used for the crystal growth. According to the desired stoichiometry (Ba$_{1-x}$Na$_x$):Fe:As were used in a molar ratio of 1:4:4. The corresponding Na was used in its metallic form and was placed at the bottom of the alumina crucible, while the well ground mixture of the pre-reacted pnictide material was carefully placed on top. The alumina crucible was then put into a niobium container which was sealed under 0.5~atm pressure of Argon in arc-melting facility. The sealed niobium crucible assembly was placed in vertical furnace with Ar atmosphere. The furnace was heated up to 1373~K with a rate of 100~K/h where it remained for 10~hours to ensure homogenous melting and afterwards cooled down to 1023~K with a rate of 2~K/h. Finally the furnace was cooled to room temperature with 300~K/hour. Platelet-like single crystals of cm-size were obtained as demonstrated by Fig.~\ref{Fig_1}(A,D). The surfaces of the crystals are shiny and metallic like (see Fig.~\ref{Fig_1}(A,D)). All crystals show layered morphology (Fig. \ref{Fig_1}(d-f)). Typically, they are easy to cleave along the $ab$ plane. These cleaving planes are shown in Fig. \ref{Fig_1}(b,e).

The quality of the grown single crystals was investigated by complementary techniques. Several samples were examined with a scanning electron microscope (SEM Philips XL 30) equipped with an electron microprobe analyzer for semi-quantitative elemental analysis using the energy dispersive x-ray (EDX) mode. The composition and in particular the Na-doping level was obtained by averaging over several different points of each single crystal. The corresponding Na-doping levels are found to be $x$= 0.25, 0.35, and 0.4. Fig.~\ref{Fig_1}(b-e) exemplariliy shows the SEM images of the single crystals with Na-contents of $x$=0.25 and $x$=0.35. The layer by layer growth can be easily seen here. 

\begin{figure}
\centering
\includegraphics[width=8cm]{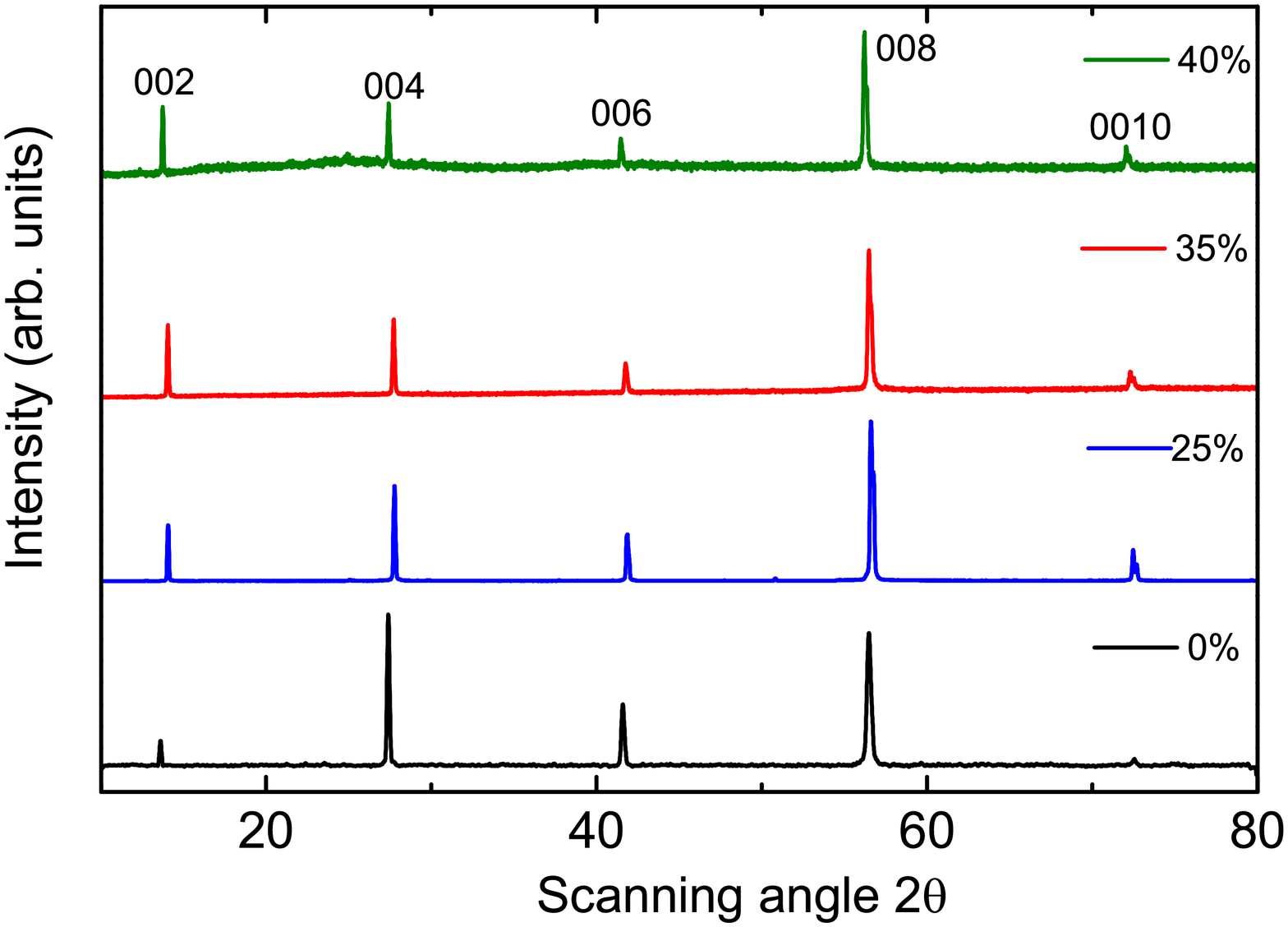}
\caption{XRD pattern of Ba$_{1-x}$Na$_x$Fe$_2$As$_2$ showing only 00$l$ reflections. The XRD data were collected using the plate-like crystals.}
\label{Fig_2}
\end{figure}

Fig.~\ref{Fig_2} shows an x-ray diffraction pattern taken on plate-like single crystals with different Na-contents using a Rigaku miniflex with Cu K$_\alpha$ radiation. Only reflections with 00\textit{l} Miller indices are observed, as expected for $c$-axis orientation. All the reflections are indexed based on the ThCr$_2$Si$_2$ structure-type, which confirms the phase purity of our Ba$_{1-x}$Na$_x$Fe$_2$As$_2$ single crystals.

Temperature dependent electronic transport was measured using a standard 4-probe alternating current dc-method with the current applied parallel to the $ab$-plane. Electrical contacts parallel to the $ab$-plane were made using thin copper wires attached to the sample with silver epoxy.  

Magnetization measurements were performed using either the Magnetic Properties Measurement System (MPMS) or the Vibrating Sample Magnetometer (VSM) from Quantum Design. A physical property measurement system (PPMS) with magnetic fields up to 9~T was used to perform the specific heat measurements. During the heat capacity measurements, the sample was cooled to the lowest temperature with an applied magnetic field (fc) and the specific heat data were collected between 1.8~K and 200~K (upon warming) by using the relaxation time method. 
Photoemission experiments for this study have been carried out using synchrotron radiation from the BESSY storage ring at the '1-cubed ARPES'  station  equipped with a $^3$He cryostat. Each sample was cleaved directly in the UHV and it was ensured that the samples exhibited a mirror-like surface before the spectra were recorded.

\section{Results and discussion}
\subsection{Magnetic measurements}
\label{mag}
\begin{figure}[ht]
\centering
\includegraphics[width=8cm]{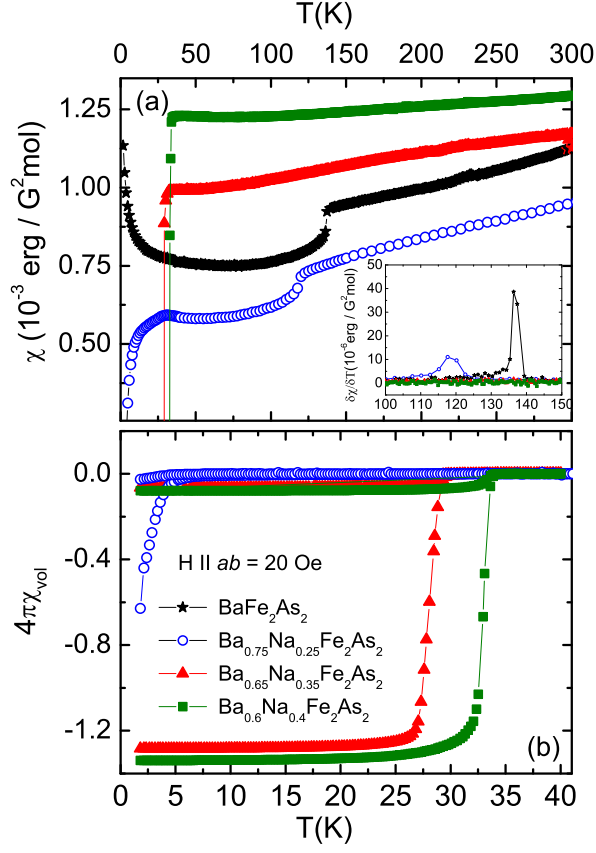}
\caption{(a) Susceptibility $\chi$ = M/B for Ba$_{1-x}$Na$_x$Fe$_2$As$_2$ for different doping levels at H$\parallel$$ab$ = 1~T. The inset shows the derivative of the static susceptibility. (b) Temperature dependence of the volume susceptibilty $\chi_{vol}$ following zfc-fc protocol as described in the text, all data have been collected for B$\parallel$ $ab$ and the applied field was 20~Oe.}
\label{Fig_3}
\end{figure}

Magnetization measurements have been performed after cooling the samples in zero (zfc) and applied (fc) magnetic field from far above the critical temperature.
Fig.~\ref{Fig_3}(a) shows the molar susceptibility $(\chi_{mol})$ measured in an applied magnetic field of 1~T with H$\parallel$$ab$. The susceptibility data of the parent compound clearly shows an anamoly at 137~K which corresponds to the combined structural ($T_s$) and magnetic ($T_{SDW}$) transition in BaFe$_2$As$_2$ (see e.g.\ \cite{ANFL11}). This transition is clearly suppressed and shifted to lower temperatures upon the substitution of Ba by Na. Specifically, the $T_s$/$T_{SDW}$ transition for Ba$_{0.75}$Na$_{0.25}$Fe$_2$As$_2$ occurs at 123~K. Moreover, this transition is also significantly broadened compared to the parent compound. The shift and broadening of the $T_s$/$T_{SDW}$ transition is more apparent in the derivative of the static susceptibility (inset of fig. \ref{Fig_3}(a)). No indication of splitting of the structural and magnetic transition is observed. However, in case of a hole-doped Ba$_{0.86}$K$_{0.14}$Fe$_2$As$_2$ crystal in the underdoped regime grown from Sn flux, a clear splitting of $T_{S}$ and $T_{SDW}$, monitored by distinct anomalies in both d$\rho$/dT and specific heat, was reported by Urbano $et$ $al.$ \cite{UGM10}. In contrast to that Avci $et$ $al.$ found no splitting at all doping levels of polycrystalline Ba$_{1-x}$K$_{x}$Fe$_2$As$_2$ where $T_s$ and $T_{SDW}$ co-exist \cite{ACG11}. Generally, samples grown from self-flux are considered to have a higher quality than those grown from Sn flux. As an example, Mathieu $et$ $al.$ could show that Sn may even substitute Ba in BaFe$_2$As$_2$ \cite{ML09}. 
Such difference in quality related to the different flux types might explain the absence/presence of the $T_{S}$ and $T_{SDW}$ splitting. At T=10~K , Ba$_{0.75}$Na$_{0.25}$Fe$_2$As$_2$ undergoes also a superconducting transition, which is a general feature of underdoped samples. Samples with higher Na-contents ($x$=0.35, 0.4) show no indication neither for a structural nor a magnetic transition, but a superconducting transition only.

Interestingly, all samples show a linear temperature dependence of the susceptibility $\chi$(T) (see fig. \ref{Fig_3}). The range of this linearity is from 300~K to about 150~K for samples exhibiting a structural and magnetic transition ($x$=0, 0.25) and 300~K to 50~K for samples with $x$=0.35, 0.4. This linearity of the susceptibility has already been discussed for undoped BaFe$_2$As$_2$ and both electron-doped LaO$_{1-x}$F$_x$FeAs and Ca(Fe$_{1-x}$Co$_x$)$_2$As$_2$ \cite{KLH10,WWW09}. So it was concluded that this linearity is a general feature and that the normal state susceptibility is not affected by the type of charge carriers. In the following, we will analyse the slope of the linear part of the magnetic susceptibility. The average slope of the susceptibility $\text{d} \chi / \text{d} T$ at high temperature is 8 $\times$10$^{-7}$~erg/(G$^{2}$mol K) which is similiar to the value of the slope reported for the electron-doped LaO$_{1-x}$F$_x$FeAs and Ca(Fe$_{1-x}$Co$_x$)$_2$As$_2$ \cite{KLH10}. However, the slope of the parent and underdoped compound is very similar (1$\times$10$^{-7}$~erg/(G$^{2}$mol K)) and decreases significantly for the samples with higher Na-contents (8$\times$10$^{-7}$~erg/(G$^{2}$molK)  for $x$=0.35; 3$\times$10$^{-7}$~erg/(G$^{2}$molK) for $x$=0.4). One may conclude that the linearity in susceptibility is a characteristic property of Fe-based superconductors, but that the slope might differ for different types of charge carriers. 

Fig.~\ref{Fig_3}(b) presents the temperature dependent volume susceptibility $(\chi_{vol})$. $\chi$ has been deduced from the measured magnetization and is not corrected for demagnetization effects. The sharp superconducting transition with a transition width of less than 2.5~K confirms the good quality of our crystals. We determine $T_c$ from the bifurcation point between the zfc and fc magnetization. Using this approach, we estimate $T_c$ to be 10~K, 29~K and 34~K for samples with $x$ =  0.25, 0.35 and 0.4, respectively. The critical temperature of our single crystals are in good agreement with the ones observed for polycrystalline samples of Na-doped BaFe$_2$As$_2$ by Cortes-Gil $et$ $al.$ \cite{CRD10}.

\subsection{Electronic Transport}
\subsubsection{Resistivity}
\label{tr}

\begin{figure}[ht]
\centering
\includegraphics[width=9cm]{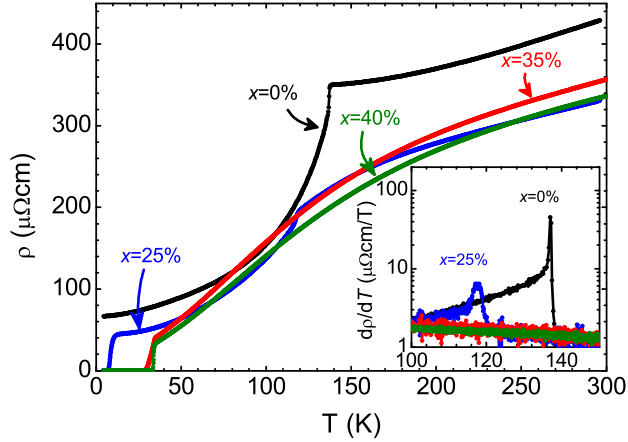}
\caption{In-plane electronic resistivity $\rho$ of Ba$_{1-x}$Na$_x$Fe$_2$As$_2$ single crystals for different Na-contents $x$ in dependence of the temperature. Inset: Derivative $\text{d}\rho/\text{d}T$ around the combined structural and magnetic transition.}
\label{Fig_4}
\end{figure}

Fig.~\ref{Fig_4} shows the resistivity of all Ba$_{1-x}$Na$_x$Fe$_2$As$_2$ single crystals.
For the undoped and $x$ = 0.25 samples the derivative of the resistivity (inset fig.~\ref{Fig_4}) shows a clear maximum at $T^{0.0}$=137~K and $T^{0.25}$=117~K which results in an upward jump of the resistivity, typically found in in hole-doped 122-compounds (see e.g. \cite{shen2011}) and is an indication for the structural and magnetic transition. Note that we did not see any indication of a doping induced splitting of this transition as observed in electron-doped 122-compounds \cite{fang2009,rullier2009,ANFL11}.  The normal state resistivity decreases by 20\% with doping which is also observed in K-doped Ba122-compounds \cite{shen2011} in contrast to a decrease of $50\,\%$ for electron-doped samples \cite{fang2009,rullier2009,ANFL11}. The sample with ($x$ = 0.25) shows both an antiferromagnetic and a superconducting transition which is typically seen in all underdoped 122-compounds \cite{NTY08,ANFL11,HSF11}. In general, a transition towards superconductivity could be observed in all samples containing Na. The critical temperature was assigned where the resistivity drops to 50\% of its value in normal state. Using this approach, superconductivity is found below $T_c^{0.25}$ = 9~K, $T_c^{0.35}$ = 32~K $T_c^{0.40}$ = 34~K, which is in line with the $T_c$ derived from magnetization measurements.

\subsubsection{Hall Effect}
\begin{figure}[ht]
\centering
\includegraphics[width=9cm]{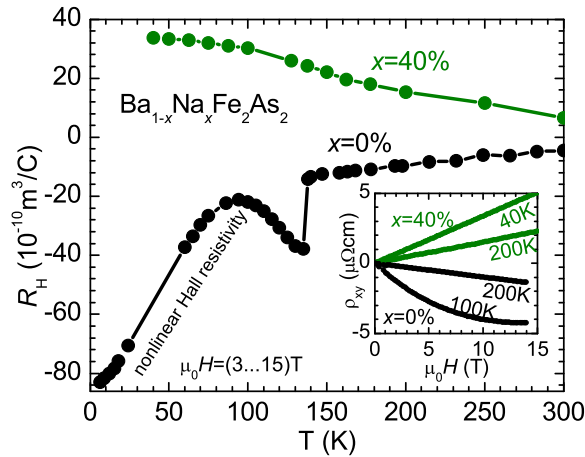}
\caption{Hall coefficient $R_H$ of optimally doped Ba$_{0.6}$Na$_{0.4}$Fe$_2$As$_2$ and undoped compound in dependence of temperature. Inset: Hall resistivity in dependence of applied magnetic field $\mu_{0} H$ for both samples at different temperatures. R$_H$ was determined from the average slope $\rho_{x,y}$ between 3~T and 15~T.}
\label{Fig_5}
\end{figure}

The Hall effect was measured for the parent and optimally doped compounds ($x$ = 0, 0.4) using the same technique as for the resistivity measurement but with Hall contacts attached. The magnetic field was applied perpendicular to the $ab$-plane while the current was passed along the $ab$-plane. I order to eliminate any offset of the measured Hall resistivity due to a finite misalignement of the Hall contacts, the Hall resistivity was calculated using the difference of measurements in positive and negative applied magnetic field.
The Hall coefficient $R_H$ of the undoped compound is negative in the whole temperature range (see Fig.~\ref{Fig_5}) and shows a linear decrease with decreasing temperature in the paramagnetic state. At the combined  structural and magnetic transition ($137\,\text{K}$), the Hall coefficient shows a jump-like behaviour and a strong non-linear temperature dependence in the antiferromagnetic regime. In contrast to this, the optimally doped compound with $x$ = 0.4 exhibits a positive Hall coefficient over the full temperature range with a linear increase with decreasing temperature. A similar behaviour was observed for the K-doped 122 sister compound and the positive Hall coefficient is consistent with hole-doping upon the substitution of Ba by K or, in the present case, by Na.

\subsection{Specific heat capacity Cp}

\begin{figure}[ht]
\centering
\includegraphics[width=9cm]{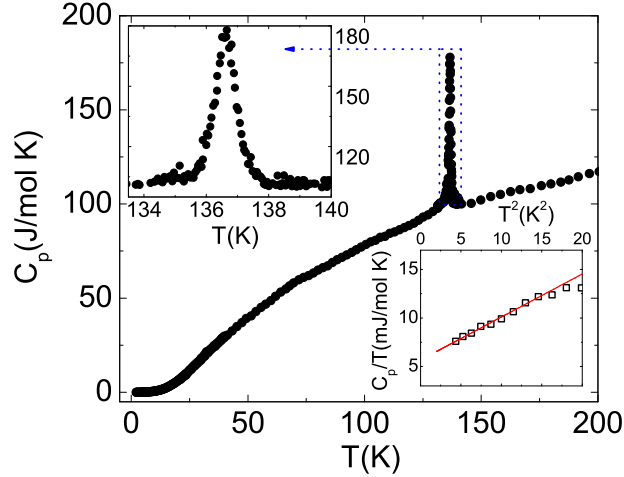}
\caption{Temperature dependence of the heat capacity measurement on a single crystal of the pristine BaFe$_{2}$As$_{2}$. The combined structural and magnetic transition transition is clearly seen at 137~K. The upper inset shows the vicinity of the phase transition on an enlarged scale, while the lower inset shows a linear fir to the data in the temperature range between 4 and 20~K.}
\label{Fig_6}
\end{figure}

Fig.~\ref{Fig_6} shows the temperature dependence of the heat capacity for a BaFe$_{2}$As$_{2}$ single crystal. The combined structural and magnetic transition occurs at 137~K with a rather sharp transition corresponding to a specific heat jump of $\Delta$C $\sim$~98~J/molK (see upper inset of Fig.\ref{Fig_6}). This transition is sharper than previously reported for BaFe$_{2}$As$_{2}$ by e.g.\ \cite{RMT08,SMJ09,KDG10} and indicates the high quality of our crystals.  
The inset of Fig.\ref{Fig_6} shows the same data near the phase transition on an enlarged scale. The full width at half maximum of the sharp peak is found to be $\sim$ 0.6~K.  The transition temperature measured by specific heat is in a good agreement with measurements of the combined $T_s$/$T_{SDW}$ transition by means of magnetization and transport (see section~\ref{tr} and section~\ref{mag}) and with earlier reports (see e.g. \cite{RMT08,KDG10}). The low temperature specific heat data of BaFe$_{2}$As$_{2}$ can be fitted linearly in the range between 4 and 20~K to C$_{p}$ = $\gamma$$T$ + $\beta$T$^{3}$ (see lower inset of Fig.\ref{Fig_6}), where $\gamma$ and $\beta$ are the electronic and lattice coefficients of the specific heat, respectively. The analysis of the data yields a Sommerfeld coefficient $\gamma$ of 6.13(8)~mJ/mol K$^{2}$ and a Debye temperature of $\Theta$$_{D}$~ = 297~K. The value for both $\gamma$ and the Debye temperature are comparable with those reported by Sefat $et$ $al.$ \cite{SMJ09}.

\begin{figure}[ht]
\centering
\includegraphics[width=8cm]{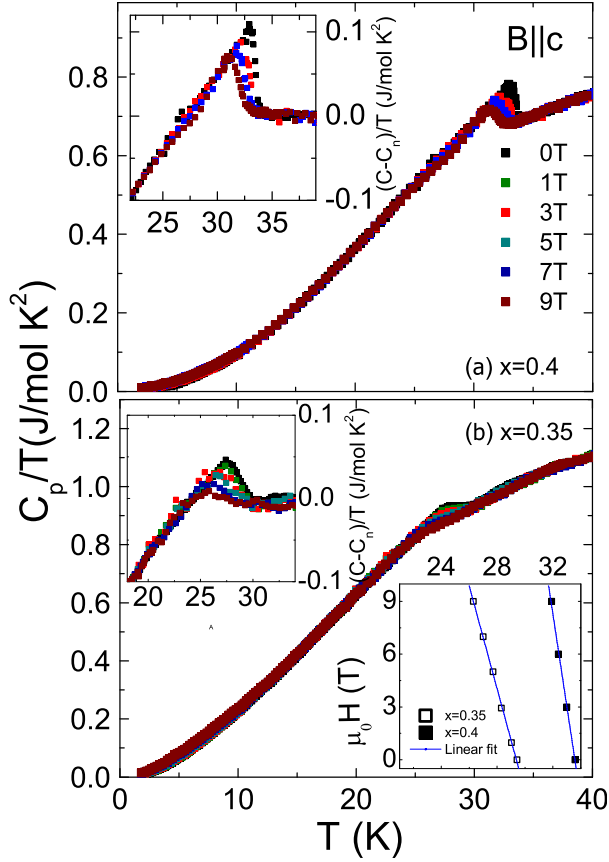}
\caption{The temperature dependent specific heat of (a) Ba$_{0.6}$Na$_{0.4}$Fe$_2$As$_2$ (b) Ba$_{0.65}$Na$_{0.35}$Fe$_2$As$_2$ in various applied magnetic fields up to 9~T parallel to the $c$-axis. The top inset shows a zoom into the superconducting state for the $x$=0.4 crystal. The lower inset shows the phase diagram of Hc$_{2}$ versus temperature and the blue dotted line represents the linear fit.}
\label{Fig_7}
\end{figure}

Fig.~\ref{Fig_7} shows the temperature dependent specific heat measurements for Ba$_{0.6}$Na$_{0.4}$Fe$_2$As$_2$ (a) and Ba$_{0.65}$Na$_{0.35}$Fe$_2$As$_2$ (b) single crystals. The measurements prove the bulk nature of superconductivity with a very sharp specific heat jump at the superconducting transition temperature $T_c$=29~K and $T_c$=34~K for $x$ =0.35 and 0.4, respectively. The critical temperatures as derived from the specific heat data (see below) are in agreement with the $T_c$ found by resistivity and magnetization measurements (see section~\ref{mag} and section~\ref{tr}). The inset of Fig.~\ref{Fig_7}(a) shows the superconducting state of the $x$=0.4 crystal in different magnetic fields with $B$ parallel to the $c$-axis. Note that these curves were obtained after subtracting the electronic $C_{el}$ and phononic $C_{ph}$ contribution to the specific heat following eq.\ref{eq1} (see \cite{ZLW11}).

\begin{equation}
C_{n} = C_{el} + C_{ph}  = \gamma T + \beta T^{3} + \eta T^{5.}
\label{eq1}
\end{equation}
At low temperature i.e.\, the specific heat is described by the sum of both contributions  $C_{n} = C_{el} + C_{ph}  = \gamma T + \beta T^{3}$, however, the term $\eta$T$^{5}$ may be added to improve the fit at higher temperatures. Using eq.~\ref{eq1} to subtract the electronic $C_{el}$ and phononic $C_{ph}$ contribution to the specific heat, the specific heat jump of the zero-field measurements $\Delta$C/$T_c$ near the transition is found to be $\sim$ 102 mJ/mol K$^{2}$ for the $x$=0.4 sample. Our $\Delta$C/$T_c$ value is comparable to the $\Delta$C/$T_c$ reported for a single crystal of the sister compound Ba$_{0.6}$K$_{0.4}$Fe$_2$As$_2$ \cite{Welp2009}. $\Delta$C/$T_c$ for the $x$=0.35 sample amounts to $\sim$ 73 mJ/mol K$^{2}$. Further details about the specific heat jump in Ba$_{0.65}$Na$_{0.35}$Fe$_2$As$_2$ and about the superconduting gap properties are reported in \cite{PAA11}. 

Next, we will explore the magnetic phase diagram for the superconducting samples with $x$ = 0.35, 0.40. Upon applying various fields with B$\parallel$c, the superconducting transition is systematically shifted to lower temperature by about 1.5~K, reduced in height and broadened as shown in the insets of Fig.~\ref{Fig_7}(a,b). 

The exact value of the transition temperature $T_c$ is determined using an ´entropy conserving construction´ for each field \cite{GTF89}. To extract the upper critical field Hc$_{2}$(0) we used the single-band Werthamer-Helfand-Hohenberg (WHH) model \cite{WHH66} H$_{c2}$(0)= - 0.69 $T_{c}$(dH$_{c2}$/dT)$_{Tc}$ as shown by the blue dotted line in the lower inset of Fig.~\ref{Fig_7}(b).  The magnetic phase diagram of both samples is perfectly described by a linear fit with average slopes of  -(dH$_{c2}$/dT)$_{Tc}$ = 3.3~T/K and -(dH$_{c2}$/dT)$_{Tc}$ =~5.25 T/K for $x$ = 0.35 and 0.40, respectively. From these values, the upper critical field H$^c_{c2}$(0) is found to be 66~T and 121~T for $x$=0.35 and 0.40, respectively. The value for the upper critical field of the $x$=0.4 compound is comparable with the critical field reported for Ba$_{0.6}$K$_{0.4}$Fe$_2$As$_2$ \cite{Welp2009}. In summary, our specific heat analysis underlines the similarities of the Na-doped compounds to the K-doped sister compounds.

\subsection{ARPES}
\begin{figure}[ht]
\centering
\includegraphics[width=16cm]{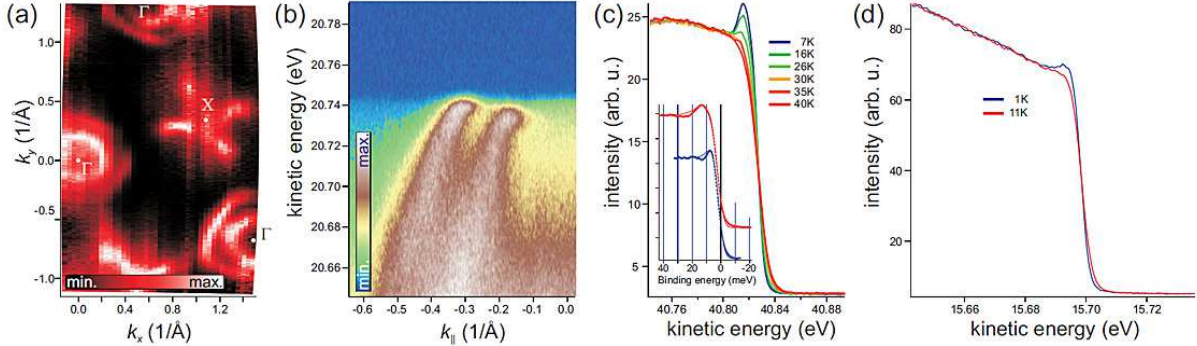}
\caption{(a) Fermi surface (FS) map, recorded at 80~eV excitation energy from the Ba$_{1-x}$Na$_x$Fe$_2$As$_2$  samples with $x$=0.35 ($T_{c}$=29K). The photoemission intensity distribution was integrated within $\pm7$meV around the Fermi level. (b) An energy-momentum cut, passing through the $\Gamma$ point and imaging the inner and outer $\Gamma$-barrels, recorded at T=1~K for the sample with $x$=0.4 ($T_{c}$=34~K). A smaller superconducting gap can be observed for the outer $\Gamma$-barrel and a larger for the inner $\Gamma$-barrel. (c) Temperature dependence of the density of states, related to the inner $\Gamma$-barrel, recorded for the sample with $x$=0.4.  Inset: Density of states for outer and inner $\Gamma$-barrels at 7~K, fitted to Dynes function \cite{EIZ09}. The determined values for the superconducting gap are $3\pm0.5$meV and $10.5\pm1$meV. (d) Density of states for the inner $\Gamma$-barrel, recorded at 1 and 11~K for the underdoped sample with $x$=0.25 ($T_{c}$=10~K). The value of the superconducting gap, estimated from the fit, is $1.9\pm0.5$meV.}
\label{Fig_8}
\end{figure}

Angle-resolved photoemission spectroscopy (ARPES) measurements were carried out on freshly cleaved surfaces of Ba$_{1-x}$Na$_x$Fe$_2$As$_2$  samples with $x$ = 0.25, 0.35, 0.4. 
Fig.~\ref{Fig_8}(a) exemplarily shows a typical photoemission intensity distribution at the Fermi level, a so-called Fermi surface (FS) map, of the Ba$_{1-x}$Na$_x$Fe$_2$As$_2$  sample with $x$=0.35. In general, the FS of all Ba$_{1-x}$Na$_x$Fe$_2$As$_2$ samples consists of hole-like sheets at the center of the Brillouin zone (BZ) ($\Gamma$ point), and of a propeller-like structure at the BZ corner (X point) with hole-like propeller blades and small electron-like FS sheets in the center. Thus, the FS of Na-substituted BaFe$_{2}$As$_{2}$ is to a high extent the same as the FS found for K-substituted BaFe$_{2}$As$_{2}$ \cite{EIZ09,ZIE08}, suggesting that the substitution of Ba by either Na or K affects the electronic band dispersion at the Fermi level in a very similar way. 

An energy-momentum cut, passing through the $\Gamma$ point (see Fig.~\ref{Fig_8}(b)), reveals not only well-defined band dispersions for the inner and outer $\Gamma$-barrels, but also the superconducting gap with a larger magnitude for the inner barrels and with a smaller magnitude for the outer one. The analysis of the temperature dependence of the electronic spectrum of the sample with $x$=0.4 ($T_{c}$=34~K) for different locations in the momentum space (see Fig.~\ref{Fig_8}(c)), allows us to conclude that the superconducting gap distribution is very close to the one observed for the optimally doped Ba$_{1-x}$K$_x$Fe$_2$As$_2$ \cite{EIZ09}: the gap magnitude maximizes for the inner $\Gamma$-barrel (around 10.5~meV) and minimizes for the outer $\Gamma$-barrel (around 3~meV). For the underdoped sample with $x$=0.25 (T$_{c}$=10~K) the opening of the superconducting gap was also monitored by the growth of the coherence peak in the density of states below $T_{c}$ (Fig.~\ref{Fig_8}(d)). The estimated gap value for the inner $\Gamma$-barrel in this case is around 2~meV.

\section{Conclusions and Summary}

In summary, single crystals of Ba$_{1-x}$Na$_x$Fe$_2$As$_2$ with $x$ = 0, 0.25, 0.35, 0.40 were grown using a self-flux method. The superconducting and normal state properties have been systematically studied by means of temperature dependent magnetic susceptibility, electrical resistivity, Hall cofficients ($R_H$) and specific heat measurements. Substitution of Ba by Na leads to the suppression of SDW ordering and induces superconductivity up to 34~K for $x$=0.4. The positive Hall coefficient of Ba$_{0.6}$Na$_{0.4}$Fe$_2$As$_2$ confirms that Na-substitution results in hole-doping similar to K-doping in Ba$_{1-x}$K$_x$Fe$_2$As$_2$. Even though the properties of our single crystals do not differ from the previously reported polycrystals, we open now way to study also the electronic structure of another type of hole-doped BaFe$_{2}$As$_{2}$. The investigation of the electronic structure of our Ba$_{1-x}$Na$_x$Fe$_2$As$_2$ single crystals by ARPES reveal striking similarities of the Fermi surface with the famous K- (hole)-doped sister compounds. Our results suggest a generic behaviour of the 122 series upon hole-doping.

\section{Acknowledgements}
The authors thank M.~Deutschmann, S.~Pichl and S.~Gass for technical support. Work was supported by
the Deutsche Forschungsgemeinschaft DFG through the Priority Programme SPP1458 (Grants No. BE1749/13 and No. GR3330/2). SW acknowledges support by DFG under the Emmy-Noether program (Grant No. WU595/3-1).


\begin{thebibliography}{33}
\expandafter\ifx\csname natexlab\endcsname\relax\def\natexlab#1{#1}\fi
\expandafter\ifx\csname bibnamefont\endcsname\relax
  \def\bibnamefont#1{#1}\fi
\expandafter\ifx\csname bibfnamefont\endcsname\relax
  \def\bibfnamefont#1{#1}\fi
\expandafter\ifx\csname citenamefont\endcsname\relax
  \def\citenamefont#1{#1}\fi
\expandafter\ifx\csname url\endcsname\relax
  \def\url#1{\texttt{#1}}\fi
\expandafter\ifx\csname urlprefix\endcsname\relax\def\urlprefix{URL }\fi
\providecommand{\bibinfo}[2]{#2}
\providecommand{\eprint}[2][]{\url{#2}}

\bibitem[{\citenamefont{Kamihara et~al.}(2008)\citenamefont{Kamihara, Watanabe,
  Hirano, and Hosono}}]{KWH08}
\bibinfo{author}{\bibfnamefont{Y.}~\bibnamefont{Kamihara}},
  \bibinfo{author}{\bibfnamefont{T.}~\bibnamefont{Watanabe}},
  \bibinfo{author}{\bibfnamefont{M.}~\bibnamefont{Hirano}}, \bibnamefont{and}
  \bibinfo{author}{\bibfnamefont{H.}~\bibnamefont{Hosono}},
  \bibinfo{journal}{J. Am. Chem. Soc.} \textbf{\bibinfo{volume}{130}},
  \bibinfo{pages}{3296} (\bibinfo{year}{2008}).

\bibitem[{\citenamefont{Ren et~al.}(2008)\citenamefont{Ren, Lu, Yang, Yi, Shen,
  Li, Che, Dong, Sun, Zhou et~al.}}]{RLY08}
\bibinfo{author}{\bibfnamefont{Z.~A.} \bibnamefont{Ren}},
  \bibinfo{author}{\bibfnamefont{W.}~\bibnamefont{Lu}},
  \bibinfo{author}{\bibfnamefont{J.}~\bibnamefont{Yang}},
  \bibinfo{author}{\bibfnamefont{W.}~\bibnamefont{Yi}},
  \bibinfo{author}{\bibfnamefont{X.-L.} \bibnamefont{Shen}},
  \bibinfo{author}{\bibfnamefont{Z.~C.} \bibnamefont{Li}},
  \bibinfo{author}{\bibfnamefont{G.~C.} \bibnamefont{Che}},
  \bibinfo{author}{\bibfnamefont{X.~L.} \bibnamefont{Dong}},
  \bibinfo{author}{\bibfnamefont{L.~L.} \bibnamefont{Sun}},
  \bibinfo{author}{\bibfnamefont{F.}~\bibnamefont{Zhou}}, \bibnamefont{et~al.},
  \bibinfo{journal}{Chin. Phys. Lett.} \textbf{\bibinfo{volume}{25}},
  \bibinfo{pages}{2215} (\bibinfo{year}{2008}).

\bibitem[{\citenamefont{Tegel et~al.}(2008)\citenamefont{Tegel, Johansson,
  Weiss, Schellenberg, Hermes, P\"ottgen, and Johrendt}}]{TJW08}
\bibinfo{author}{\bibfnamefont{M.}~\bibnamefont{Tegel}},
  \bibinfo{author}{\bibfnamefont{S.}~\bibnamefont{Johansson}},
  \bibinfo{author}{\bibfnamefont{V.}~\bibnamefont{Weiss}},
  \bibinfo{author}{\bibfnamefont{I.}~\bibnamefont{Schellenberg}},
  \bibinfo{author}{\bibfnamefont{W.}~\bibnamefont{Hermes}},
  \bibinfo{author}{\bibfnamefont{R.}~\bibnamefont{P\"ottgen}},
  \bibnamefont{and} \bibinfo{author}{\bibfnamefont{D.}~\bibnamefont{Johrendt}},
  \bibinfo{journal}{Europhys. Lett.} \textbf{\bibinfo{volume}{84}},
  \bibinfo{pages}{67007} (\bibinfo{year}{2008}).

\bibitem[{\citenamefont{Rotter et~al.}(2008{\natexlab{a}})\citenamefont{Rotter,
  Tegel, and Johrendt}}]{RTJ08}
\bibinfo{author}{\bibfnamefont{M.}~\bibnamefont{Rotter}},
  \bibinfo{author}{\bibfnamefont{M.}~\bibnamefont{Tegel}}, \bibnamefont{and}
  \bibinfo{author}{\bibfnamefont{D.}~\bibnamefont{Johrendt}},
  \bibinfo{journal}{Phys. Rev. Lett.} \textbf{\bibinfo{volume}{101}},
  \bibinfo{pages}{107006} (\bibinfo{year}{2008}{\natexlab{a}}).

\bibitem[{\citenamefont{Sasmal et~al.}(2008)\citenamefont{Sasmal, Lv, Lorenz,
  Guloy, Chen, Xue, and Chu}}]{Sll08}
\bibinfo{author}{\bibfnamefont{K.}~\bibnamefont{Sasmal}},
  \bibinfo{author}{\bibfnamefont{B.}~\bibnamefont{Lv}},
  \bibinfo{author}{\bibfnamefont{B.}~\bibnamefont{Lorenz}},
  \bibinfo{author}{\bibfnamefont{A.~M.} \bibnamefont{Guloy}},
  \bibinfo{author}{\bibfnamefont{F.}~\bibnamefont{Chen}},
  \bibinfo{author}{\bibfnamefont{Y.-Y.} \bibnamefont{Xue}}, \bibnamefont{and}
  \bibinfo{author}{\bibfnamefont{C.-W.} \bibnamefont{Chu}},
  \bibinfo{journal}{Phys. Rev. Lett.} \textbf{\bibinfo{volume}{101}},
  \bibinfo{pages}{107007} (\bibinfo{year}{2008}).

\bibitem[{\citenamefont{Tapp et~al.}(2008)\citenamefont{Tapp, Tang, Lv, Sasmal,
  Lorenz, Chu, and Guloy}}]{TTL08}
\bibinfo{author}{\bibfnamefont{J.~H.} \bibnamefont{Tapp}},
  \bibinfo{author}{\bibfnamefont{Z.}~\bibnamefont{Tang}},
  \bibinfo{author}{\bibfnamefont{B.}~\bibnamefont{Lv}},
  \bibinfo{author}{\bibfnamefont{K.}~\bibnamefont{Sasmal}},
  \bibinfo{author}{\bibfnamefont{B.}~\bibnamefont{Lorenz}},
  \bibinfo{author}{\bibfnamefont{P.~C.~W.} \bibnamefont{Chu}},
  \bibnamefont{and} \bibinfo{author}{\bibfnamefont{A.~M.} \bibnamefont{Guloy}},
  \bibinfo{journal}{Phys. Rev. B} \textbf{\bibinfo{volume}{78}},
  \bibinfo{pages}{060505} (\bibinfo{year}{2008}).

\bibitem[{\citenamefont{Hsu et~al.}(2008)\citenamefont{Hsu, Luo, Yeh, Chen,
  Huang, Wu, Lee, Huang, Chu, Yan et~al.}}]{HLK08}
\bibinfo{author}{\bibfnamefont{F.~C.} \bibnamefont{Hsu}},
  \bibinfo{author}{\bibfnamefont{J.~Y.} \bibnamefont{Luo}},
  \bibinfo{author}{\bibfnamefont{K.~W.} \bibnamefont{Yeh}},
  \bibinfo{author}{\bibfnamefont{T.~K.} \bibnamefont{Chen}},
  \bibinfo{author}{\bibfnamefont{T.~W.} \bibnamefont{Huang}},
  \bibinfo{author}{\bibfnamefont{P.~M.} \bibnamefont{Wu}},
  \bibinfo{author}{\bibfnamefont{Y.~C.} \bibnamefont{Lee}},
  \bibinfo{author}{\bibfnamefont{Y.~L.} \bibnamefont{Huang}},
  \bibinfo{author}{\bibfnamefont{Y.~Y.} \bibnamefont{Chu}},
  \bibinfo{author}{\bibfnamefont{D.~C.} \bibnamefont{Yan}},
  \bibnamefont{et~al.}, \bibinfo{journal}{Proceedings of the National Academy
  of Sciences} \textbf{\bibinfo{volume}{105}}, \bibinfo{pages}{14262}
  (\bibinfo{year}{2008}).

\bibitem[{\citenamefont{Zhigadlo et~al.}(2008)\citenamefont{Zhigadlo, Katrych,
  Bukowski, Weyeneth, Puzniak, and Karpinski.}}]{ZKB08}
\bibinfo{author}{\bibfnamefont{N.~D.} \bibnamefont{Zhigadlo}},
  \bibinfo{author}{\bibfnamefont{S.}~\bibnamefont{Katrych}},
  \bibinfo{author}{\bibfnamefont{Z.}~\bibnamefont{Bukowski}},
  \bibinfo{author}{\bibfnamefont{S.}~\bibnamefont{Weyeneth}},
  \bibinfo{author}{\bibfnamefont{R.}~\bibnamefont{Puzniak}}, \bibnamefont{and}
  \bibinfo{author}{\bibfnamefont{J.}~\bibnamefont{Karpinski.}},
  \bibinfo{journal}{J. Phys.: Condens. Matter} \textbf{\bibinfo{volume}{20}},
  \bibinfo{pages}{342202} (\bibinfo{year}{2008}).

\bibitem[{\citenamefont{Z.~Ban}(1965)}]{BS65}
\bibinfo{author}{\bibfnamefont{M.~S.} \bibnamefont{Z.~Ban}},
  \bibinfo{journal}{Acta Cryst.} \textbf{\bibinfo{volume}{18}},
  \bibinfo{pages}{594} (\bibinfo{year}{1965}).

\bibitem[{VC9(1991)}]{VC91}
\emph{\bibinfo{title}{Pearson´s Handbook of Crystallographic Data for
  Intermetallic Phases, second ed. ASM International}} (\bibinfo{publisher}{P.
  Villars, L. D. Calvert}, \bibinfo{address}{Materials Park, OH},
  \bibinfo{year}{1991}).

\bibitem[{\citenamefont{Colombier et~al.}(2009)\citenamefont{Colombier, Bud'ko,
  Ni, and Canfield}}]{CBNC09}
\bibinfo{author}{\bibfnamefont{E.}~\bibnamefont{Colombier}},
  \bibinfo{author}{\bibfnamefont{S.~L.} \bibnamefont{Bud'ko}},
  \bibinfo{author}{\bibfnamefont{N.}~\bibnamefont{Ni}}, \bibnamefont{and}
  \bibinfo{author}{\bibfnamefont{P.~C.} \bibnamefont{Canfield}},
  \bibinfo{journal}{Phys. Rev. B} \textbf{\bibinfo{volume}{79}},
  \bibinfo{pages}{224518} (\bibinfo{year}{2009}).

\bibitem[{\citenamefont{Cortes-Gil et~al.}(2010)\citenamefont{Cortes-Gil,
  Parker, Pitcher, Hadermann, and Clarke}}]{CRD10}
\bibinfo{author}{\bibfnamefont{R.}~\bibnamefont{Cortes-Gil}},
  \bibinfo{author}{\bibfnamefont{D.~R.} \bibnamefont{Parker}},
  \bibinfo{author}{\bibfnamefont{M.~J.} \bibnamefont{Pitcher}},
  \bibinfo{author}{\bibfnamefont{J.}~\bibnamefont{Hadermann}},
  \bibnamefont{and} \bibinfo{author}{\bibfnamefont{S.~J.}
  \bibnamefont{Clarke}}, \bibinfo{journal}{Chemistry of Materials}
  \textbf{\bibinfo{volume}{22}}, \bibinfo{pages}{4304} (\bibinfo{year}{2010}).

\bibitem[{\citenamefont{Aswartham et~al.}(2011)\citenamefont{Aswartham, Nacke,
  Friemel, Leps, Wurmehl, Wizent, Hess, Klingeler, Behr, Singh
  et~al.}}]{ANFL11}
\bibinfo{author}{\bibfnamefont{S.}~\bibnamefont{Aswartham}},
  \bibinfo{author}{\bibfnamefont{C.}~\bibnamefont{Nacke}},
  \bibinfo{author}{\bibfnamefont{G.}~\bibnamefont{Friemel}},
  \bibinfo{author}{\bibfnamefont{N.}~\bibnamefont{Leps}},
  \bibinfo{author}{\bibfnamefont{S.}~\bibnamefont{Wurmehl}},
  \bibinfo{author}{\bibfnamefont{N.}~\bibnamefont{Wizent}},
  \bibinfo{author}{\bibfnamefont{C.}~\bibnamefont{Hess}},
  \bibinfo{author}{\bibfnamefont{R.}~\bibnamefont{Klingeler}},
  \bibinfo{author}{\bibfnamefont{G.}~\bibnamefont{Behr}},
  \bibinfo{author}{\bibfnamefont{S.}~\bibnamefont{Singh}},
  \bibnamefont{et~al.}, \bibinfo{journal}{J. Cryst. Growth}
  \textbf{\bibinfo{volume}{314}}, \bibinfo{pages}{341} (\bibinfo{year}{2011}).

\bibitem[{\citenamefont{Urbano et~al.}(2010)\citenamefont{Urbano, Green,
  Moulton, Reyes, Kuhns, Bittar, Adriano, Garitezi, Bufaical, and
  Pagliuso}}]{UGM10}
\bibinfo{author}{\bibfnamefont{R.~R.} \bibnamefont{Urbano}},
  \bibinfo{author}{\bibfnamefont{E.~L.} \bibnamefont{Green}},
  \bibinfo{author}{\bibfnamefont{W.~G.} \bibnamefont{Moulton}},
  \bibinfo{author}{\bibfnamefont{A.~P.} \bibnamefont{Reyes}},
  \bibinfo{author}{\bibfnamefont{P.~L.} \bibnamefont{Kuhns}},
  \bibinfo{author}{\bibfnamefont{E.~M.} \bibnamefont{Bittar}},
  \bibinfo{author}{\bibfnamefont{C.}~\bibnamefont{Adriano}},
  \bibinfo{author}{\bibfnamefont{T.~M.} \bibnamefont{Garitezi}},
  \bibinfo{author}{\bibfnamefont{L.}~\bibnamefont{Bufaical}}, \bibnamefont{and}
  \bibinfo{author}{\bibfnamefont{P.~G.} \bibnamefont{Pagliuso}},
  \bibinfo{journal}{Phys. Rev. L} \textbf{\bibinfo{volume}{105}},
  \bibinfo{pages}{107001} (\bibinfo{year}{2010}).

\bibitem[{\citenamefont{Avci et~al.}(2011)\citenamefont{Avci, Chmaissem,
  Goremychkin, Rosenkranz, Castellan, Chung, Todorov, Schlueter, Claus,
  Kanatzidis et~al.}}]{ACG11}
\bibinfo{author}{\bibfnamefont{S.}~\bibnamefont{Avci}},
  \bibinfo{author}{\bibfnamefont{O.}~\bibnamefont{Chmaissem}},
  \bibinfo{author}{\bibfnamefont{E.~A.} \bibnamefont{Goremychkin}},
  \bibinfo{author}{\bibfnamefont{S.}~\bibnamefont{Rosenkranz}},
  \bibinfo{author}{\bibfnamefont{J.-P.} \bibnamefont{Castellan}},
  \bibinfo{author}{\bibfnamefont{D.~Y.} \bibnamefont{Chung}},
  \bibinfo{author}{\bibfnamefont{I.~S.} \bibnamefont{Todorov}},
  \bibinfo{author}{\bibfnamefont{J.~A.} \bibnamefont{Schlueter}},
  \bibinfo{author}{\bibfnamefont{H.}~\bibnamefont{Claus}},
  \bibinfo{author}{\bibfnamefont{M.~G.} \bibnamefont{Kanatzidis}},
  \bibnamefont{et~al.}, \bibinfo{journal}{Phys. Rev. B}
  \textbf{\bibinfo{volume}{83}}, \bibinfo{pages}{172503}
  (\bibinfo{year}{2011}).

\bibitem[{\citenamefont{Mathieu and Latturner}(2009)}]{ML09}
\bibinfo{author}{\bibfnamefont{J.~L.} \bibnamefont{Mathieu}} \bibnamefont{and}
  \bibinfo{author}{\bibfnamefont{S.~E.} \bibnamefont{Latturner}},
  \bibinfo{journal}{Chem. Commun.} p. \bibinfo{pages}{4965}
  (\bibinfo{year}{2009}).

\bibitem[{\citenamefont{Klingeler et~al.}(2010)\citenamefont{Klingeler, Leps,
  Hellmann, Popa, Stockert, Hess, Kataev, Grafe, Hammerath, Lang
  et~al.}}]{KLH10}
\bibinfo{author}{\bibfnamefont{R.}~\bibnamefont{Klingeler}},
  \bibinfo{author}{\bibfnamefont{N.}~\bibnamefont{Leps}},
  \bibinfo{author}{\bibfnamefont{I.}~\bibnamefont{Hellmann}},
  \bibinfo{author}{\bibfnamefont{A.}~\bibnamefont{Popa}},
  \bibinfo{author}{\bibfnamefont{U.}~\bibnamefont{Stockert}},
  \bibinfo{author}{\bibfnamefont{C.}~\bibnamefont{Hess}},
  \bibinfo{author}{\bibfnamefont{V.}~\bibnamefont{Kataev}},
  \bibinfo{author}{\bibfnamefont{H.-J.} \bibnamefont{Grafe}},
  \bibinfo{author}{\bibfnamefont{F.}~\bibnamefont{Hammerath}},
  \bibinfo{author}{\bibfnamefont{G.}~\bibnamefont{Lang}}, \bibnamefont{et~al.},
  \bibinfo{journal}{Phys. Rev. B} \textbf{\bibinfo{volume}{81}},
  \bibinfo{pages}{024506} (\bibinfo{year}{2010}).

\bibitem[{\citenamefont{Wang et~al.}(2009)\citenamefont{Wang, Wu, Wu, Chen,
  Xie, Ying, Yan, Liu, and Chen}}]{WWW09}
\bibinfo{author}{\bibfnamefont{X.~F.} \bibnamefont{Wang}},
  \bibinfo{author}{\bibfnamefont{T.}~\bibnamefont{Wu}},
  \bibinfo{author}{\bibfnamefont{G.}~\bibnamefont{Wu}},
  \bibinfo{author}{\bibfnamefont{H.}~\bibnamefont{Chen}},
  \bibinfo{author}{\bibfnamefont{Y.~L.} \bibnamefont{Xie}},
  \bibinfo{author}{\bibfnamefont{J.~J.} \bibnamefont{Ying}},
  \bibinfo{author}{\bibfnamefont{Y.~J.} \bibnamefont{Yan}},
  \bibinfo{author}{\bibfnamefont{R.~H.} \bibnamefont{Liu}}, \bibnamefont{and}
  \bibinfo{author}{\bibfnamefont{X.~H.} \bibnamefont{Chen}},
  \bibinfo{journal}{Phys. Rev. Lett.} \textbf{\bibinfo{volume}{102}},
  \bibinfo{pages}{117005} (\bibinfo{year}{2009}).

\bibitem[{\citenamefont{Shen et~al.}(2011)\citenamefont{Shen, Yang, Wang, Han,
  Zeng, Shan, Ren, and Wen}}]{shen2011}
\bibinfo{author}{\bibfnamefont{B.}~\bibnamefont{Shen}},
  \bibinfo{author}{\bibfnamefont{H.}~\bibnamefont{Yang}},
  \bibinfo{author}{\bibfnamefont{Z.-S.} \bibnamefont{Wang}},
  \bibinfo{author}{\bibfnamefont{F.}~\bibnamefont{Han}},
  \bibinfo{author}{\bibfnamefont{B.}~\bibnamefont{Zeng}},
  \bibinfo{author}{\bibfnamefont{L.}~\bibnamefont{Shan}},
  \bibinfo{author}{\bibfnamefont{C.}~\bibnamefont{Ren}}, \bibnamefont{and}
  \bibinfo{author}{\bibfnamefont{H.-H.} \bibnamefont{Wen}},
  \bibinfo{journal}{Phys. Rev. B} \textbf{\bibinfo{volume}{84}},
  \bibinfo{pages}{184512} (\bibinfo{year}{2011}).

\bibitem[{\citenamefont{Fang et~al.}(2009)\citenamefont{Fang, Luo, Cheng, Wang,
  Jia, Mu, Shen, Mazin, Shan, Ren et~al.}}]{fang2009}
\bibinfo{author}{\bibfnamefont{L.}~\bibnamefont{Fang}},
  \bibinfo{author}{\bibfnamefont{H.}~\bibnamefont{Luo}},
  \bibinfo{author}{\bibfnamefont{P.}~\bibnamefont{Cheng}},
  \bibinfo{author}{\bibfnamefont{Z.}~\bibnamefont{Wang}},
  \bibinfo{author}{\bibfnamefont{Y.}~\bibnamefont{Jia}},
  \bibinfo{author}{\bibfnamefont{G.}~\bibnamefont{Mu}},
  \bibinfo{author}{\bibfnamefont{B.}~\bibnamefont{Shen}},
  \bibinfo{author}{\bibfnamefont{I.~I.} \bibnamefont{Mazin}},
  \bibinfo{author}{\bibfnamefont{L.}~\bibnamefont{Shan}},
  \bibinfo{author}{\bibfnamefont{C.}~\bibnamefont{Ren}}, \bibnamefont{et~al.},
  \bibinfo{journal}{Phys. Rev. B} \textbf{\bibinfo{volume}{80}},
  \bibinfo{pages}{140508} (\bibinfo{year}{2009}).

\bibitem[{\citenamefont{Rullier-Albenque
  et~al.}(2009)\citenamefont{Rullier-Albenque, Colson, and
  Alloul}}]{rullier2009}
\bibinfo{author}{\bibfnamefont{F.}~\bibnamefont{Rullier-Albenque}},
  \bibinfo{author}{\bibfnamefont{D.}~\bibnamefont{Colson}}, \bibnamefont{and}
  \bibinfo{author}{\bibfnamefont{H.}~\bibnamefont{Alloul}},
  \bibinfo{journal}{Phys. Rev. Lett.} \textbf{\bibinfo{volume}{103}},
  \bibinfo{pages}{057001} (\bibinfo{year}{2009}).

\bibitem[{\citenamefont{Ni et~al.}(2008)\citenamefont{Ni, Tillman, Yan,
  Kracher, Hannahs, Bud'ko, and Canfield}}]{NTY08}
\bibinfo{author}{\bibfnamefont{N.}~\bibnamefont{Ni}},
  \bibinfo{author}{\bibfnamefont{M.~E.} \bibnamefont{Tillman}},
  \bibinfo{author}{\bibfnamefont{J.-Q.} \bibnamefont{Yan}},
  \bibinfo{author}{\bibfnamefont{A.}~\bibnamefont{Kracher}},
  \bibinfo{author}{\bibfnamefont{S.~T.} \bibnamefont{Hannahs}},
  \bibinfo{author}{\bibfnamefont{S.~L.} \bibnamefont{Bud'ko}},
  \bibnamefont{and} \bibinfo{author}{\bibfnamefont{P.~C.}
  \bibnamefont{Canfield}}, \bibinfo{journal}{Phys. Rev. B}
  \textbf{\bibinfo{volume}{78}}, \bibinfo{pages}{214515}
  (\bibinfo{year}{2008}).

\bibitem[{\citenamefont{Harnagea et~al.}(2011)\citenamefont{Harnagea, Singh,
  Friemel, Leps, Bombor, Abdel-Hafiez, Wolter, Hess, Klingeler, Behr
  et~al.}}]{HSF11}
\bibinfo{author}{\bibfnamefont{L.}~\bibnamefont{Harnagea}},
  \bibinfo{author}{\bibfnamefont{S.}~\bibnamefont{Singh}},
  \bibinfo{author}{\bibfnamefont{G.}~\bibnamefont{Friemel}},
  \bibinfo{author}{\bibfnamefont{N.}~\bibnamefont{Leps}},
  \bibinfo{author}{\bibfnamefont{D.}~\bibnamefont{Bombor}},
  \bibinfo{author}{\bibfnamefont{M.}~\bibnamefont{Abdel-Hafiez}},
  \bibinfo{author}{\bibfnamefont{A.~U.~B.} \bibnamefont{Wolter}},
  \bibinfo{author}{\bibfnamefont{C.}~\bibnamefont{Hess}},
  \bibinfo{author}{\bibfnamefont{R.}~\bibnamefont{Klingeler}},
  \bibinfo{author}{\bibfnamefont{G.}~\bibnamefont{Behr}}, \bibnamefont{et~al.},
  \bibinfo{journal}{Phys. Rev. B} \textbf{\bibinfo{volume}{83}},
  \bibinfo{pages}{094523} (\bibinfo{year}{2011}).

\bibitem[{\citenamefont{Rotter et~al.}(2008{\natexlab{b}})\citenamefont{Rotter,
  Tegel, Johrendt, Schellenberg, Hermes, and P\"ottgen}}]{RMT08}
\bibinfo{author}{\bibfnamefont{M.}~\bibnamefont{Rotter}},
  \bibinfo{author}{\bibfnamefont{M.}~\bibnamefont{Tegel}},
  \bibinfo{author}{\bibfnamefont{D.}~\bibnamefont{Johrendt}},
  \bibinfo{author}{\bibfnamefont{I.}~\bibnamefont{Schellenberg}},
  \bibinfo{author}{\bibfnamefont{W.}~\bibnamefont{Hermes}}, \bibnamefont{and}
  \bibinfo{author}{\bibfnamefont{R.}~\bibnamefont{P\"ottgen}},
  \bibinfo{journal}{Phys. Rev. B} \textbf{\bibinfo{volume}{78}},
  \bibinfo{pages}{020503} (\bibinfo{year}{2008}{\natexlab{b}}).

\bibitem[{\citenamefont{Sefat et~al.}(2009)\citenamefont{Sefat, McGuire, Jin,
  Sales, Mandrus, Ronning, Bauer, and Mozharivskyj}}]{SMJ09}
\bibinfo{author}{\bibfnamefont{A.~S.} \bibnamefont{Sefat}},
  \bibinfo{author}{\bibfnamefont{M.~A.} \bibnamefont{McGuire}},
  \bibinfo{author}{\bibfnamefont{R.}~\bibnamefont{Jin}},
  \bibinfo{author}{\bibfnamefont{B.~C.} \bibnamefont{Sales}},
  \bibinfo{author}{\bibfnamefont{D.}~\bibnamefont{Mandrus}},
  \bibinfo{author}{\bibfnamefont{F.}~\bibnamefont{Ronning}},
  \bibinfo{author}{\bibfnamefont{E.~D.} \bibnamefont{Bauer}}, \bibnamefont{and}
  \bibinfo{author}{\bibfnamefont{Y.}~\bibnamefont{Mozharivskyj}},
  \bibinfo{journal}{Phys. Rev. B} \textbf{\bibinfo{volume}{79}},
  \bibinfo{pages}{094508} (\bibinfo{year}{2009}).

\bibitem[{\citenamefont{Kant et~al.}(2010)\citenamefont{Kant, Deisenhofer,
  G\"unther, Schrettle, Loidl, Rotter, and Johrendt}}]{KDG10}
\bibinfo{author}{\bibfnamefont{C.}~\bibnamefont{Kant}},
  \bibinfo{author}{\bibfnamefont{J.}~\bibnamefont{Deisenhofer}},
  \bibinfo{author}{\bibfnamefont{A.}~\bibnamefont{G\"unther}},
  \bibinfo{author}{\bibfnamefont{F.}~\bibnamefont{Schrettle}},
  \bibinfo{author}{\bibfnamefont{A.}~\bibnamefont{Loidl}},
  \bibinfo{author}{\bibfnamefont{M.}~\bibnamefont{Rotter}}, \bibnamefont{and}
  \bibinfo{author}{\bibfnamefont{D.}~\bibnamefont{Johrendt}},
  \bibinfo{journal}{Phys. Rev. B} \textbf{\bibinfo{volume}{81}},
  \bibinfo{pages}{014529} (\bibinfo{year}{2010}).

\bibitem[{\citenamefont{Zhao et~al.}(2011)\citenamefont{Zhao, Liu, Wang, Deng,
  Lv, Zhu, Li, and Jin}}]{ZLW11}
\bibinfo{author}{\bibfnamefont{K.}~\bibnamefont{Zhao}},
  \bibinfo{author}{\bibfnamefont{Q.~Q.} \bibnamefont{Liu}},
  \bibinfo{author}{\bibfnamefont{X.~C.} \bibnamefont{Wang}},
  \bibinfo{author}{\bibfnamefont{Z.}~\bibnamefont{Deng}},
  \bibinfo{author}{\bibfnamefont{Y.~X.} \bibnamefont{Lv}},
  \bibinfo{author}{\bibfnamefont{J.~L.} \bibnamefont{Zhu}},
  \bibinfo{author}{\bibfnamefont{F.~Y.} \bibnamefont{Li}}, \bibnamefont{and}
  \bibinfo{author}{\bibfnamefont{C.~Q.} \bibnamefont{Jin}},
  \bibinfo{journal}{Phys. Rev. B} \textbf{\bibinfo{volume}{84}},
  \bibinfo{pages}{184534} (\bibinfo{year}{2011}).

\bibitem[{\citenamefont{Welp et~al.}(2009)\citenamefont{Welp, Xie, Koshelev,
  Kwok, Luo, Wang, Mu, and Wen}}]{Welp2009}
\bibinfo{author}{\bibfnamefont{U.}~\bibnamefont{Welp}},
  \bibinfo{author}{\bibfnamefont{R.}~\bibnamefont{Xie}},
  \bibinfo{author}{\bibfnamefont{A.~E.} \bibnamefont{Koshelev}},
  \bibinfo{author}{\bibfnamefont{W.~K.} \bibnamefont{Kwok}},
  \bibinfo{author}{\bibfnamefont{H.~Q.} \bibnamefont{Luo}},
  \bibinfo{author}{\bibfnamefont{Z.~S.} \bibnamefont{Wang}},
  \bibinfo{author}{\bibfnamefont{G.}~\bibnamefont{Mu}}, \bibnamefont{and}
  \bibinfo{author}{\bibfnamefont{H.~H.} \bibnamefont{Wen}},
  \bibinfo{journal}{Phys. Rev. B} \textbf{\bibinfo{volume}{79}},
  \bibinfo{pages}{094505} (\bibinfo{year}{2009}).

\bibitem[{\citenamefont{Pramanik et~al.}(2011)\citenamefont{Pramanik,
  Abdel-Hafiez, Aswartham, Wolter, Wurmehl, Kataev, and B\"uchner}}]{PAA11}
\bibinfo{author}{\bibfnamefont{A.~K.} \bibnamefont{Pramanik}},
  \bibinfo{author}{\bibfnamefont{M.}~\bibnamefont{Abdel-Hafiez}},
  \bibinfo{author}{\bibfnamefont{S.}~\bibnamefont{Aswartham}},
  \bibinfo{author}{\bibfnamefont{A.~U.~B.} \bibnamefont{Wolter}},
  \bibinfo{author}{\bibfnamefont{S.}~\bibnamefont{Wurmehl}},
  \bibinfo{author}{\bibfnamefont{V.}~\bibnamefont{Kataev}}, \bibnamefont{and}
  \bibinfo{author}{\bibfnamefont{B.}~\bibnamefont{B\"uchner}},
  \bibinfo{journal}{Phys. Rev. B} \textbf{\bibinfo{volume}{84}},
  \bibinfo{pages}{064525} (\bibinfo{year}{2011}).

\bibitem[{\citenamefont{J.E.~Gordon and Phillips}(1989)}]{GTF89}
\bibinfo{author}{\bibfnamefont{R.~F.} \bibnamefont{J.E.~Gordon},
  \bibfnamefont{M.L.~Tan}} \bibnamefont{and}
  \bibinfo{author}{\bibfnamefont{N.}~\bibnamefont{Phillips}},
  \bibinfo{journal}{Solid State Comm.} \textbf{\bibinfo{volume}{69}},
  \bibinfo{pages}{625} (\bibinfo{year}{1989}).

\bibitem[{\citenamefont{Werthamer et~al.}(1966)\citenamefont{Werthamer,
  Helfand, and Hohenberg}}]{WHH66}
\bibinfo{author}{\bibfnamefont{N.~R.} \bibnamefont{Werthamer}},
  \bibinfo{author}{\bibfnamefont{E.}~\bibnamefont{Helfand}}, \bibnamefont{and}
  \bibinfo{author}{\bibfnamefont{P.~C.} \bibnamefont{Hohenberg}},
  \bibinfo{journal}{Phys. Rev.} \textbf{\bibinfo{volume}{147}},
  \bibinfo{pages}{295} (\bibinfo{year}{1966}).

\bibitem[{\citenamefont{Evtushinsky et~al.}(2009)\citenamefont{Evtushinsky,
  Inosov, Zabolotnyy, Koitzsch, Knupfer, B\"uchner, Viazovska, Sun, Hinkov,
  Boris et~al.}}]{EIZ09}
\bibinfo{author}{\bibfnamefont{D.~V.} \bibnamefont{Evtushinsky}},
  \bibinfo{author}{\bibfnamefont{D.~S.} \bibnamefont{Inosov}},
  \bibinfo{author}{\bibfnamefont{V.~B.} \bibnamefont{Zabolotnyy}},
  \bibinfo{author}{\bibfnamefont{A.}~\bibnamefont{Koitzsch}},
  \bibinfo{author}{\bibfnamefont{M.}~\bibnamefont{Knupfer}},
  \bibinfo{author}{\bibfnamefont{B.}~\bibnamefont{B\"uchner}},
  \bibinfo{author}{\bibfnamefont{M.~S.} \bibnamefont{Viazovska}},
  \bibinfo{author}{\bibfnamefont{G.~L.} \bibnamefont{Sun}},
  \bibinfo{author}{\bibfnamefont{V.}~\bibnamefont{Hinkov}},
  \bibinfo{author}{\bibfnamefont{A.~V.} \bibnamefont{Boris}},
  \bibnamefont{et~al.}, \bibinfo{journal}{Phys. Rev. B}
  \textbf{\bibinfo{volume}{79}}, \bibinfo{pages}{054517}
  (\bibinfo{year}{2009}).

\bibitem[{\citenamefont{Zabolotnyy et~al.}(2008)\citenamefont{Zabolotnyy,
  Inosov, Evtushinsky, Koitzsch, Kordyuk, Sun, Park, Haug, Hinkov, Boris
  et~al.}}]{ZIE08}
\bibinfo{author}{\bibfnamefont{V.~B.} \bibnamefont{Zabolotnyy}},
  \bibinfo{author}{\bibfnamefont{D.~S.} \bibnamefont{Inosov}},
  \bibinfo{author}{\bibfnamefont{D.~V.} \bibnamefont{Evtushinsky}},
  \bibinfo{author}{\bibfnamefont{A.}~\bibnamefont{Koitzsch}},
  \bibinfo{author}{\bibfnamefont{A.~A.} \bibnamefont{Kordyuk}},
  \bibinfo{author}{\bibfnamefont{G.~L.} \bibnamefont{Sun}},
  \bibinfo{author}{\bibfnamefont{J.~T.} \bibnamefont{Park}},
  \bibinfo{author}{\bibfnamefont{D.}~\bibnamefont{Haug}},
  \bibinfo{author}{\bibfnamefont{V.}~\bibnamefont{Hinkov}},
  \bibinfo{author}{\bibfnamefont{A.~V.} \bibnamefont{Boris}},
  \bibnamefont{et~al.}, \bibinfo{journal}{Nature}
  \textbf{\bibinfo{volume}{457}}, \bibinfo{pages}{569} (\bibinfo{year}{2008}).

\end{thebibliography}

\end{document}